\documentstyle[11pt,amstex,amssymb,epsf]{article} 

\def\rect#1#2{{\vcenter{\vbox{\hrule height.3pt
            \hbox{\vrule width.3pt height#2truecm \kern#1truecm
            \vrule width.3pt}
            \hrule height.3pt}}}}

\newtheorem{theorem}{Theorem} 

\newtheorem{lemma}{Lemma}

\begin{document}
\begin{center}
{
\vskip1cm
{\Large 
A note on the metastability of the Ising model: the alternate 
updating case}
\vskip1.0cm
{\large Emilio N.M.\ Cirillo}
}
\vskip0.8cm
Dipartimento Me.\ Mo.\ Mat., Universit\`a degli Studi di Roma ``La Sapienza,"\\
via A. Scarpa 16, I--00161, Roma, Italy.
\end{center}

\vskip 1.5 cm 
\begin{abstract} 
We study the metastable behavior of the two-dimensional Ising model in the
case of an alternate updating rule: parallel updating of spins on the even
(odd) sublattice are permitted at even (odd) times.  We show that
although the dynamics is different from the Glauber serial case the typical
exit path from the metastable phase remains the same.
\end{abstract}


\newpage
There have been many studies of the problem of
metastability in the Ising model, i.e.\ the time evolution of a system
initially in the all minus phase in a small positive external magnetic field.
The case of Glauber-Metropolis serial dynamics has been
discussed in \cite{[NS1]}, where it was proved that the exit from the
metastable phase is achieved via the nucleation of a critical droplet and
that the life time of the metastable phase depends only on the energy of
such a critical droplet. The proof, which uses the pathwise approach
introduced in \cite{[CGOV]}, is based on the analysis of the tendency of
droplets of pluses in a background of minuses to shrink or to grow.  In
\cite{[NS2]} it was shown that the behavior of such droplets does not
change if the spin flip rates are modified.

In this note we consider the case of parallel dynamics in which spins on
the even (odd) sublattice are simultaneously updated. More precisely: we
consider the two-dimensional Ising model on a finite rectangle
$\Lambda\subset{\Bbb Z}^2$ with even side length and periodic boundary
conditions; for any $x\in\Lambda$, $\sigma(x)=\pm 1$ denotes the spin
variable at that site and to any configuration
$\sigma\in\Omega:=\{-1,+1\}^{\Lambda}$ we associate the energy
\begin{equation}
H(\sigma):=-\frac{J}{2}\sum_{<x,y>}\sigma(x)\sigma(y) 
-\frac{h}{2}\sum_{x\in\Lambda}\sigma(x)
\;\; ,
\label{eq:ham}
\end{equation}
where $J\gg h >0$.  The 
equilibrium is described by the Gibbs measure
$\nu(\sigma):=\exp\{-\beta H(\sigma)\}/Z$
with  $\beta$, the inverse of the temperature, a positive real number
and 
$Z:=\sum_{\eta\in\Omega} \exp\{-\beta H(\eta)\}$ 
the partition function.

We define the following updating rule: 
we partition $\Lambda$ into its even and odd sublattices, 
$\Lambda=\Lambda^e\cup\Lambda^o$, with 
$\Lambda^{e(o)}:=
\{x=(x_1,x_2)\in\Lambda:\; x_1+x_2\; {{\mathrm{is\; even\;(odd)}}}\}$
and 
we denote by
$\Xi^{e(o)}:=\{\{x_1,...,x_m\}:\; 1\le m\le|\Lambda|/2,\; x_i\in \Lambda^{e(o)}
               \textrm{ for all } i=1,...,m\}$ 
the set of all possible collections of sites in
$\Lambda^{e(o)}$. Given $I\in\Xi^{e(o)}$ we denote by
$\sigma^I$ the configuration obtained by flipping in $\sigma$ the
$|I|$ spins associated to the sites in $I$.  
Let $\mu^{e(o)}$ be a probability measure on $\Xi^{e(o)}$ such that 
$\mu^{e(o)}(I)>0$ for any $I\in\Xi^{e(o)}$ and $t=0,1,\dots$ an integer
variable. Finally, at any even (odd) time $t$ a set $I\in\Xi^{e(o)}$ is
chosen at random with probability $\mu^{e(o)}(I)$   
and the $|I|$ spins associated to sites in $I$ are flipped with
probability $\exp\{-\beta[H(\sigma^I)-H(\sigma)]^+\}$, where $\sigma$ is
the configuration at time $t$. In other words we can say that 
the
evolution of model (\ref{eq:ham}) is described by a Markov chain $\sigma_t$
with transition probability
$P(\sigma_{t+1}=\eta|\sigma_{t}=\xi)=p_{\alpha}(\xi,\eta)$, with
$\alpha=e,o$ depending on $t$ even or odd, where 
for any $\eta\not=\xi$ 
\begin{equation}
p_{\alpha}(\xi,\eta):=\left\{
\begin{array}{cc}
\mu^{\alpha}(I)
\exp\{-\beta [H(\xi^{I})-H(\xi)]^+\}&
{\mathrm{if}}\;\exists I\in\Xi^{\alpha}\; 
{\mathrm{such\; that}}\; \eta=\xi^{I}\\
0 &{\mathrm{otherwise}}\\
\end{array}
\right.
\;\;\; 
\label{eq:transizione}
\end{equation}
and $p_{\alpha}(\xi,\xi):=1-\sum_{\eta\not=\xi}p_{\alpha}(\xi,\eta)$. 
We remark that this dynamics satisfies the detailed balance condition
with respect to the Gibbs measure,
hence this is its unique invariant measure.

Our dynamics lives in the same configuration space with the same 
energy landscape as the standard single spin flip Metropolis algorithm, but 
many more connections between different configurations have been opened
\cite{[BCLS]}. 
In other words we allow the system to perform jumps forbidden in the 
standard serial case. To be more precise the notions of
``communicating states" (allowed jumps)
and, hence, of connected sets have been 
changed:  
we say that $\sigma,\eta\in\Omega$ are  
{\it e(o)-nearest neighbors}
(e(o)-nn) iff $\exists I\in\Xi^{e(o)}$ such that 
$\eta=\sigma^I$; two configurations are nearest neighbors if
they are e-nn or o-nn, in other words any two
configurations when differing only on the evens or odds sides are neighbors.
A set ${\cal A}\subset\Omega$ is connected iff
for any $\sigma,\eta\in{\cal A}$ there exists a sequence $\sigma_0,
...,\sigma_n$ of configuration such that: $\sigma_0=\sigma$, 
$\sigma_n=\eta$
and for any $i=0,...,n-1$, depending on $i$ even or odd,
$\sigma_i$ and $\sigma_{i+1}$ are e(o)--nn.

The fact that new jumps have been allowed could, a priori, modify the
metastable character of the dynamics and eventually destroy it. In the most
extreme case 
single spin flip paths containing steps against the energy 
drift can be replaced by direct energy favored jumps. Indeed, let us 
consider $\sigma\in\Omega$ and 
$I=\{x_i:\;1\le i\le n\}\in\Xi^{e(o)}$, it is easy to show that
\begin{equation}
H(\sigma^I)-H(\sigma)=
\sum_{i=1}^n [H(\sigma^{x_i})-H(\sigma)]
\;\;\; ; 
\label{eq:indipendenza}
\end{equation} 
one just has to remark that for any $i,j=1,...,n$ and $i\not= j$ 
the two sites $x_i$ and $x_j$ are not nearest neighbors. 
In other words (\ref{eq:indipendenza}) means that from the point of 
view of the energy a parallel event can be substituted by a 
suitable sequence of single
spin flip events. Now, one can easily choose
$\sigma$ and $I=\{x_i:\;1\le i\le n\}$ such that 
$H(\sigma^I)\le H(\sigma)$ and 
$H(\sigma^{x_i})-H(\sigma)>0$ for some $i=1,\dots,n$.
 
Nevertheless, 
by following the scheme of \cite{[OS]} and by making a repeated use of
(\ref{eq:indipendenza}) we will show that the metastable behavior of the 
system remains unchanged: the first problem to solve
is the full characterization of the local minima of the hamiltonian. A local
minimum is a configuration $\sigma\in\Omega$ such that for any nearest
neighbor configuration $\eta$ one has $H(\sigma)\le H(\eta)$. Although
the hamiltonian is the standard Ising hamiltonian, the notion of 
neighboring configurations is dramatically changed, so it is not obvious a 
priori that the local minima are the rectangles of pluses 
as in the single spin flip case. But this is the case, indeed 
we can prove that a configuration $\sigma$
is a local minimum for the alternate dynamics if and only if it  
is a local minimum in the single spin flip case. 
It is easy to see that the sufficient condition holds because a pair of
nearest neighboring configurations for the single spin flip dynamics
is a pair of nearest neighboring configuration even in the alternate case.
The necessity condition:  
let $\sigma$ be a local minimum for the single spin flip dynamics 
and let $\eta$ a e(o)-nn of $\sigma$. There exists 
$I=\{x_i:\;1\le i\le n\}\in\Xi^{e(o)}$ 
such that $\eta=\sigma^I$; the proof 
follows from (\ref{eq:indipendenza}) and from the fact that any $\sigma^{x_i}$
is a nearest neighbor of $\sigma$ in the single spin flip case.

The second step in the understanding of the metastable behavior of the model
is the description of the tendency to shrink or to grow of the local minima. 
A preliminary definition: given 
$\eta\in\Omega$ if $\sigma_t$ is the process started at some 
$\sigma_0\in\Omega$
we denote by $\tau_{\eta}:=\{t>0:\; \sigma_t=\eta\}$ the
first hitting time on $\eta$. Now 
we state the following lemma on the criticality of rectangles: a 
rectangle $R_{\ell,m}$ is a configuration with a rectangular droplet of 
pluses, with side lengths $\ell$ and $m$, plunged in the sea of minuses.

\begin{lemma}
\label{lemma:critici}
\par\noindent
Let $\ell^*:=[{2J\over h}]+1$; consider a rectangle 
$R_{\ell,m}\in\Omega$, with $\ell\le m$, and the
Markov chain $\sigma_t$ started at $\sigma_0=R_{\ell,m}$.
Given $\varepsilon>0$ we have: 
if $\ell<\ell^*$ then $R_{\ell,m}$ is subcritical, that is 
$P(\tau_{-{\underline 1}}<\tau_{+{\underline 1}}) 
\stackrel{\beta\to\infty}{\longrightarrow} 1$, and
$P(\exp\{\beta(\ell-1)h-\beta\varepsilon\}
<\tau_{-{\underline 1}}<\exp\{\beta(\ell-1)h+\beta\varepsilon\})
\stackrel{\beta\to\infty}{\longrightarrow} 1$.
If $\ell\ge \ell^*$ then $R_{\ell,m}$ is supercritical, that is 
$P(\tau_{+{\underline 1}}<\tau_{-{\underline 1}}) 
\stackrel{\beta\to\infty}{\longrightarrow} 1$, and
$P(\exp\{\beta(2J-h)-\beta\varepsilon\} 
<\tau_{+{\underline 1}}<\exp\{\beta(2J-h)+\beta\varepsilon\})
\stackrel{\beta\to\infty}{\longrightarrow} 1$.
\end{lemma} 
To prove the Lemma we
need few more definitions: a downhill path is
a sequence of configuration $\sigma_0,\,\sigma_1\, ,...,\,\sigma_n$ such
that depending on $i$ even or odd $\sigma_i$ and $\sigma_{i+1}$ are
e(o)-nn and for any $i=0,...,n-1$ $H(\sigma_i)\ge H(\sigma_{i+1})$;
an uphill path is the reverse of a downhill one.
The basin of attraction of a local minimum $\sigma$ is the set 
$B(\sigma):=\{\eta\in\Omega:\; {\mathrm{all\; the\; downhill\; paths\;
starting\; from}}\;\eta\; {\mathrm{end\; in}}\; \sigma\}$, and for any 
${\mathcal A}\subset\Omega$ its boundary $\partial{\mathcal A}$ is the 
collection of configurations
$\eta\in\Omega\setminus{\mathcal A}$ such that there exists 
$\sigma\in{\mathcal A}$ nearest neighbor of $\eta$.

Now, we just have to find the minimum of the energy on $\partial
B(R_{\ell,m})$ 
and use the results in Propositions 
3.4 and 3.7 of \cite{[OS]}. 
Starting from $R_{\ell,m}$ one should consider all the possible uphill paths
reaching $\partial B(R_{\ell,m})$, 
but 
(\ref{eq:indipendenza}) implies that all the mechanisms involving more 
than a single spin flip are energetically less favorable; 
hence the minimum of the energy on $\partial B(R_{\ell,m})$ 
can be found using the 
same arguments as in the single spin flip case.
{}From this remark the proof of the Lemma follows. 

Now we state the theorem which describes the exit from the
metastable phase: 
let us denote by ${\cal P}$ the configuration with all the spins minus
excepted those in a $\ell^*\times(\ell^*-1)$ rectangle and in a
unit protuberance attached to one of its longest sides. We set
$ \Gamma:=H({\cal P})-H(-{\underline 1})=4J\ell^*-h({\ell^*}^2-\ell^*+1)$,
we consider the process $\sigma_t$ starting from $-{\underline 1}$ 
and we define 
${\bar\tau}_{-{\underline 1}}:=\sup\{t<\tau_{+{\underline 1}}:
\; \sigma_t=-{\underline 1}\}$,  the last time the system visits 
$-{\underline 1}$ before reaching $+{\underline 1}$,   
and
${\bar\tau}_{\cal P}:=\inf\{t>{\bar\tau}_{-{\underline 1}}:\; 
\sigma_t={\cal P}\}$, the first time the system reaches ${\cal P}$ after
having ``definitively" left $-{\underline 1}$.
Finally, we can state the following
\begin{theorem}
\label{th}
\par\noindent
Let us suppose $\sigma_0=-{\underline 1}$, let $\varepsilon >0$,
then
$P({\bar\tau}_{\cal P}<\tau_{+{\underline 1}})
\stackrel{\beta\to\infty}{\longrightarrow} 1$
and
$P(\exp\{\beta\Gamma-\beta\varepsilon\}<\tau_{+{\underline 1}}
<\exp\{\beta\Gamma+\beta\varepsilon\}
\stackrel{\beta\to\infty}{\longrightarrow} 1$.
\end{theorem}
We sketch the proof of the Theorem. We use the same notation as in 
Section 3.1 of \cite{[KO]}. First of all we consider the map 
$S:\Omega\to\Omega$ morally associating to each 
configuration $\sigma\in\Omega$ the ``largest" local minimum 
$S(\sigma)$ to which $\sigma$ is connected by a 
downhill path, 
where downhill must be understood in the sense
of one spin nearest neighboring configurations (configurations differing 
for the value of a single spin). See Section 3.1 of \cite{[KO]} for
the rigorous definition of $S$. 
Now, we define the set  
${\mathcal A}\subset\Omega$ as the collection of configurations
$\sigma\in\Omega$ such that all the plus spins of $S(\sigma)$ are those
associated to the sites inside 
a collection of rectangles on the lattice such that however 
two of such rectangles are chosen, their mutual distance is greater
or equal to $\sqrt{5}$ (pairwaise non--interacting rectangles).

Following \cite{[OS]}, we prove that 
$H({\cal P})$ is the minimum of the
energy on the boundary of ${\cal A}$.
Let us partition $\partial{\mathcal A}$ into two parts 
$\partial{\mathcal A}=\partial_1{\mathcal A}
                 \cup\partial_{\ge 2}{\mathcal A}$: for each 
$\eta\in\partial{\mathcal A}$, 
$\eta\in\partial_1{\mathcal A}$ iff there exists $x\in\Lambda$ such 
that $\sigma^x\in{\mathcal A}$, while
$\eta\in\partial_{\ge 2}{\mathcal A}$ iff 
$\sigma^x\in\Omega\setminus{\mathcal A}$ for all $x\in\Lambda$ 
and there exists $I\in\Xi^{e(o)}$ with $|I|\ge 2$ such   
that $\sigma^I\in{\mathcal A}$.
We first  prove that for all $\eta\in\partial_{\ge 2}{\mathcal A}$ there exists
$\zeta\in\partial_1{\mathcal A}$ such that $H(\zeta)<H(\eta)$.
Let $\eta\in\partial_{\ge 2}{\cal A}$ and  
$I\in\Xi^{e(o)}$ the subset of $\Lambda$ with smallest cardinality such
that $\sigma:=\eta^I\in{\cal A}$. 
The definition of $\partial_{\ge 2}{\mathcal A}$ implies $|I|\ge 2$;  
the minimality of $I$ and the construction of $S$ 
imply $\eta(x)=+1$ and 
\begin{equation}
\label{aa}
H(\eta^x)<H(\eta) \textrm{ for all } x\in I
\;\;\; .
\end{equation}
Now, let $x\in I$, from the minimality of $I$ we  
get $\zeta:=\eta^{I\setminus\{x\}}\in\Omega\setminus{\mathcal A}$.
Hence, $\zeta^x=\sigma\in{\mathcal A}$ and 
$\zeta\in\Omega\setminus{\mathcal A}$
imply $\zeta\in\partial_1{\mathcal A}$. Moreover, from $|I|\ge 2$, 
equation (\ref{eq:indipendenza}) and the property (\ref{aa}) we, finally, get 
$H(\zeta)<H(\eta)$.

Now, by using the same strategy as in Section 3.1 of \cite{[KO]}
we get that the minimum of 
the energy on $\partial{\mathcal A}$ is realized in 
${\mathcal P}$. 
Finally, Theorem \ref{th} is a consequence of 
Propositions 3.4 and 3.7 of \cite{[OS]}.

In conclusion we can say that in the case of the alternate updating the 
property (\ref{eq:indipendenza}) implies that the mechanism of escape from
the metastable phase is not changed with respect to the Glauber case;  
on the other hand one can expect that things could be different if the
parallel updating were allowed all over the lattice without any restriction
\cite{[CN]}. 

\vskip 0.2 cm
\par\noindent
{\large\bf Acknowledgments}
\par\noindent 
E.C.\ wishes to express his thanks to J.L.\ Lebowitz for having 
suggested the problem and for having carefully read the manuscript.
E.C.\ also thanks F.R.\ Nardi and E.\ Olivieri for very useful discussions  
and the European  
network ``Stochastic Analysis and its Applications" ERB--FMRX--CT96--0075
for financial support.


\end{document}